Journal of Risk and Financial Management

MDPI

Article

# News-driven Expectations and Volatility Clustering

**Sabiou M. Inoua** [1,*]

[1] Chapman University
[*] Correspondence: inouasabiou@gmail.com

**Abstract:** Financial volatility obeys two fascinating empirical regularities that apply to various assets, on various markets, and on various time scales: it is fat-tailed (more precisely power-law distributed) and it tends to be clustered in time. Many interesting models have been proposed to account for these regularities, notably agent-based models, which mimic the two empirical laws through a complex mix of nonlinear mechanisms such as traders' switching between trading strategies in highly nonlinear way. This paper explains the two regularities simply in terms of traders' attitudes towards news, an explanation that follows almost by definition of the traditional dichotomy of financial market participants, investors versus speculators, whose behaviors are reduced to their simplest forms. Long-run investors' valuations of an asset are assumed to follow a news-driven random walk, thus capturing the investors' persistent, long memory of fundamental news. Short-term speculators' anticipated returns, on the other hand, are assumed to follow a news-driven autoregressive process, capturing their shorter memory of fundamental news, and, by the same token, the feedback intrinsic to the short-sighted, trend-following (or herding) mindset of speculators. These simple, linear, models of traders' expectations, it is shown, explain the two financial regularities in a generic and robust way. Rational expectations, the dominant model of traders' expectations, is not assumed here, owing to the famous no-speculation, no-trade results.

## 1. Introduction

A meticulous and extensive study of high-frequency financial data by various researchers reveals important empirical regularities. Financial volatility, in particular, obeys two well-established empirical laws that attracted special attention in the literature: it is fat-tailed (in fact power-law distributed with an exponent often close to 3) and it tends to be clustered in time, unfolding through intense bursts of high instability interrupting calmer periods (Mandelbrot, 1963; Fama, 1963; Ding, Granger, & Engle, 1993; Gopikrishnan, Meyer, Amaral, & Stanley, 1998; Lux, 1998; Plerou, Gabaix, Stanley, & Gopikrishnan, 2006; Cont, 2007; Bouchaud, 2011). The first regularity implies that extreme price changes are much more likely than suggests the standard assumption of normal distribution. The second property, volatility clustering, reveals a nontrivial predictability in the return process, whose sign is uncorrelated but whose amplitude is long-range correlated. These are fascinating regularities that apply to various financial products (commodities, stocks, indices, exchange rates, CDS[1]) on various markets and on various time scales.

The universality and robustness of these laws (illustrated graphically in section 2) suggests that there must be some *basic, permanent, and general* mechanisms causing them (an intuition that shall be the heuristic and guiding principle throughout this paper). To identify these causes requires to go back to the basics of financial theory and to contrast the two major paradigms on financial fluctuations. The dominant view today, the efficient market hypothesis, treats an asset's price as following a random walk *exogenously driven by fundamental news* (Bachelier, 1900; Osborne, 1959; Fama, 1963; Cootner, 1964; Fama, 1965a, 1965b; Samuelson, 1965; Fama, 1970). On the other hand is the growing resurgence of an old view of financial markets that insists on *endogenous amplifying feedback mechanisms* caused by mimetic or trend-following speculative expectations, fueled by credit, and responsible for bubbles and crashes (Fisher, 1933; Keynes, 1936; Shiller, 1980; Smith, Suchanek, & Williams, 1988; Cutler, Poterba, & Summers, 1989; Orléan, 1989; Cutler, Poterba, & Summers, 1990; Minsky, 1992; Caginalp, Porter, & Smith, 2000; Barberis & Thaler, 2003; Porter & Smith, 2003; Akerlof & Shiller, 2010; Shaikh, 2010; Bouchaud, 2011; Dickhaut, Lin, Porter, & Smith, 2012; Keen, 2013; Palan, 2013; Soros, 2013; Gjerstad & Smith, 2014; Soros, 2015).[2]

---

[1] See, e.g., Bouchaud and Challet (2016).

The endogenous cause of financial volatility, probably predominant in empirical data (Bouchaud, 2011), is particularly taken seriously in agent-based models, which, unlike neoclassical finance, deal explicitly with the traditional dichotomy of financial participants, investors versus speculators (often named differently), extending it to include other types of players; besides traders' heterogeneity, these models also insist on traders' learning, adaptation, interaction, etc. They generate realistic fat-tailed and clustered volatility, but typically through a *complex mix of nonlinear mechanisms*, notably traders' switching between trading strategies. These interesting models of financial volatility have already been carefully reviewed elsewhere (Cont, 2007; Samanidou, Zschischang, Stauffer, & Lux, 2007; He, Li, & Wang, 2016; Lux & Alfarano, 2016). The realism of these models comes at a price, however; for it is not easy to isolate basic causes of the financial regularities amid a mathematically intractable complex of highly nonlinear processes at work simultaneously[3]. So, while this literature contributed significantly to a faithful picture of financial markets, it is not completely satisfactory for a basic reason: the sophisticated trading behaviors commonly assumed in this literature, handled through modern computers, is hardly a natural explanation of the financial regularities, whose discovery (let alone validity) goes back to the 1960s, an early and more rudimentary stage of finance. There must be, in other words, something of a most fundamental nature, some permanent cause intrinsic to the very act of financial trading, that is causing these regularities. GARCH models are perhaps more popular and more parsimonious models of the two regularities than agent-based models (Engle, 1982; Bollerslev, 1986; Bollerslev, Chou, & Kroner, 1992); but these statistical models are hardly a theoretical explanation of the empirical laws from explicit economic mechanisms; when fitted to empirical data, moreover, they imply an infinite-variance return process, the integrated GARCH (or IGARCH) model (Engle & Bollerslev, 1986), which corresponds to a more extreme randomness than the empirical one (Mikosch & Starica, 2000, 2003).

This paper explains the two regularities simply in terms of traders' attitudes towards news, an explanation that follows almost by definition of the traditional dichotomy of financial market participants, investors versus speculators, whose behaviors are reduced to their simplest forms. Long-run investors' valuations of an asset are assumed to follow a news-driven random walk, thus capturing the investors' persistent, long memory of fundamental news. Short-term speculators' anticipated returns, on the other hand, are assumed to follow a news-driven autoregressive process, capturing their shorter memory of news, and, by the same token, the feedback intrinsic to their short-memory, trend-following (or herding) mindset. These simple, linear, models of traders' expectations, it is shown below, explain the two financial regularities in a generic and robust way. Rational expectations, the dominant model of traders' expectations, is not assumed here, owing to the famous no-speculation, no-trade results (Milgrom & Stokey, 1982; Tirole, 1982). In fact there seems to be an intrinsic difficulty in building a realistic theory of high-frequency volatility of financial markets, caused by incessant trading at almost all time scales and often driven by short-term speculative gains, from rational expectations, since they typically lead to a no-speculation, no-trade equilibrium.

---

[2] The nuance in this diverse literature on endogenous financial instability, already clearly articulated by the classical economists (Inoua & Smith, 2020), lies perhaps in the nature of the ultimate destabilizing force that is specifically emphasized in each tradition, notably human psychology (Keynes and behavioral finance) or the easy bank-issued liquidity that backs or fuels the speculative euphoria and without which this latter would be of no significant, macroeconomic, harm (Fisher, Minsky, Kindleberger, etc., and the classical economists who preceded them).

[3] Other types of models are also suggested for the power law more specifically; one of them, e.g., relates the power law of return to the trades of very large institutional investors (Plerou et al., 2006).

The model this paper suggests can be viewed as a *simple theory* of the interplay between the exogenous and endogenous causes of financial volatility, and, by the same token, identifies the two components to be responsible for the two regularities, reducing them to basic, *linear mechanisms*; it is a synthesis of the two paradigms above-mentioned, avoiding the caveats on both sides: the no-trade problem inherent to the neoclassical formulation of the news-driven random walk model, and the nonlinear complexity characteristic of agent-based models. The power-law tail of volatility can be shown to derive intrinsically from the self-reinforcing amplifications inherent to herding or trend-following speculative trading. Trend-following speculation, for example, which is a popular financial practice, leads directly to a random-coefficient autoregressive (RCAR) return process in a competitive financial market, assuming a simple *linear* competitive price adjustment, as recent empirical evidence suggests (Cont, Kukanov, & Stoikov, 2014). The RCAR model derives naturally, provided that trend-following is modeled, not in terms of moving averages of past prices (as often assumed in the agent-based literature) but in terms of *moving averages of past returns*, which is more natural and more convenient (Beekhuizen & Hallerbach, 2017). The power-law tail of such processes is rigorously proven in the mathematical literature[4] (Kesten, 1973; Klüppelberg & Pergamenchtchikov, 2004; Buraczewski, Damek, & Mikosch, 2016). But it can be proven that the RCAR model, briefly derived below[5] (in section 3), cannot explain volatility clustering (Mikosch & Starica, 2000; Basrak, Davis, & Mikosch, 2002; Mikosch & Starica, 2003; Buraczewski et al., 2016).[6] The basic cause of clustered volatility, this paper suggests, is none other than the impact of exogenous news on expectations. A more general model is therefore suggested that includes, as usual, a second class of agents besides speculators: fundamental-value investors, who attach a real value to an asset and buy it when they think the asset is underpriced, or sell it, otherwise, updating additively their valuations with the advent of a fundamental, exogenous news; the amount of information a news reveal to the traders about the asset's worth can be precisely quantified by the log-probability of the news, as is known from information theory (Shannon, 1948). Speculators' expectations are more subtle, since they are at least partly endogenous. The simplest compromise consists of modeling the speculators' anticipated return as a first-order autoregressive process with a coefficient that is lower than 1, to capture speculative self-reinforcing feedback, but close enough to 1, so that exogenous news have a persistent enough impact on speculators' expectations as well. This extended model is superior, not only by generating both the fat-tailed and clustered volatility, but also by ensuring stationarity of the return process.

---

[4] RCAR processes are also known as Kesten processes, named after H. Kesten whose seminal theorem proves their power-law tail behavior. Kesten's theorem is perhaps first used in finance to study GARCH processes, which are in fact also Kesten processes. 'Rational bubbles' also can be interpreted as first-order RCAR processes assuming a random discount factor (Lux & Sornette, 2002); but this model generates a tail exponent smaller than 1. First-order RCAR processes have also been suggested as approximations to complex agent-based mechanisms (Sato & Takayasu, 1998; Carvalho, 2001; Aoki, 2002). In this paper, however, a general RCAR return process holds directly in a competitive market of trend-following speculators.

[5] A more detailed study of the power-law tail of volatility as it emerges from the RCAR model is the subject of a planned follow-up paper.

[6] It may seem paradoxical that volatility in GARCH models, being also RCAR processes, could nonetheless generate clustered volatility; in fact, it is more precisely the square root of a GARCH process that is an RCAR process; moreover the famous so-called 'GARCH effect' is again an 'IGARCH effect' (Mikosch & Starica, 2000, 2003). It is more the persistence (hence also the near-nonstationarity) implied by the near-random walk of an IGARCH volatility that mimics the volatility clusters in these models. This near integration of the IGARCH model resembles the near integration of speculators' anticipated return in this paper's model.

## 2. The empirical regularities

Let $P_t$ be the price of a financial asset at the closing of period $t$, let the return (or relative price change) be $r_t = (P_t - P_{t-1})/P_{t-1}$. The two empirical regularities read formally: (1) $\mathbb{P}(|r_t|>x) \sim Cx^{-\alpha}$, for big values $x$, where often $\alpha \approx 3$ ($C$ being merely a normalizing constant); (2) $\mathrm{cor}(|r_t|,|r_{t+h}|) > 0$ over a long range of lags $h > 0$, while $\mathrm{cor}(r_t, r_{t+h}) \approx 0$ for all $h > 0$. Figures 1 and 2 illustrate these two regularities for General Electric's daily stock price and the NYSE daily index[7].

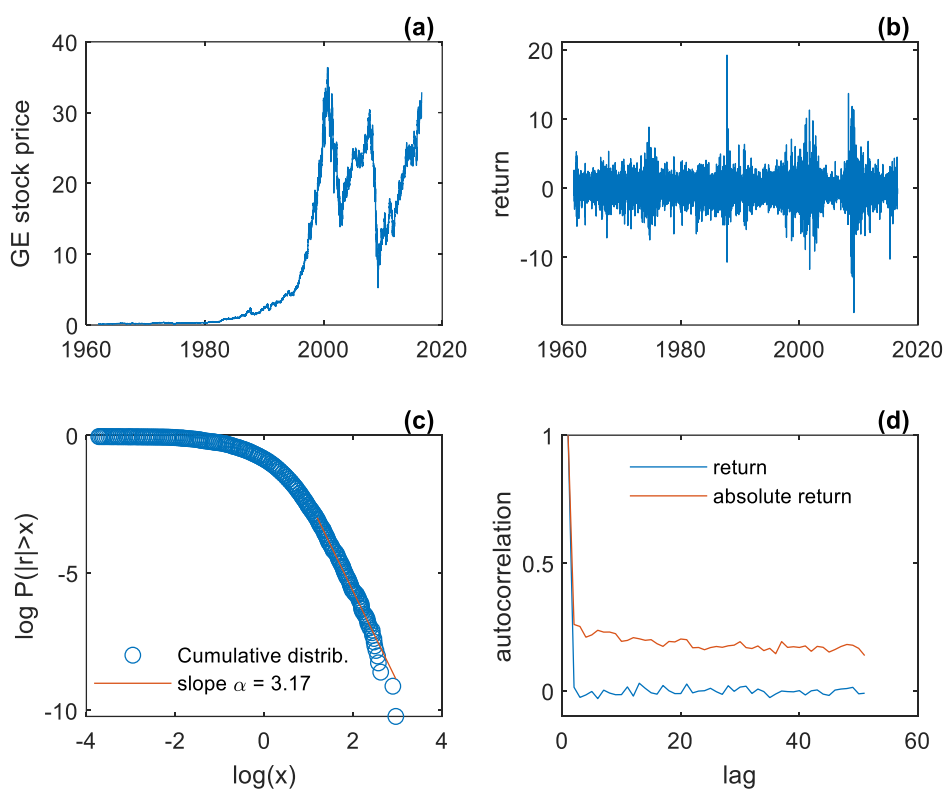

**Figure 1.** General Electric stock: (**a**) Price; (**b**) Return (in percent); (**c**) cumulative distribution of volatility in log-log scale, and a linear fit of the tail, with a slope close to 3; (**d**) Autocorrelation function of return, which is almost zero at all lags, while that of volatility is nonzero over a long range of lags (a phenomenon known as volatility clustering).

---

[7] The linear fit is based on a maximum-likelihood algorithm developed by Clauset, Shalizi, and Newman (2009), which is an important reference for the statistical test of empirical power laws; the program codes are available at http://tuvalu.santafe.edu/~aaronc/powerlaws/. For an introduction to power laws more generally, see, for example, Newman (2005) and Gabaix (2008, 2016).

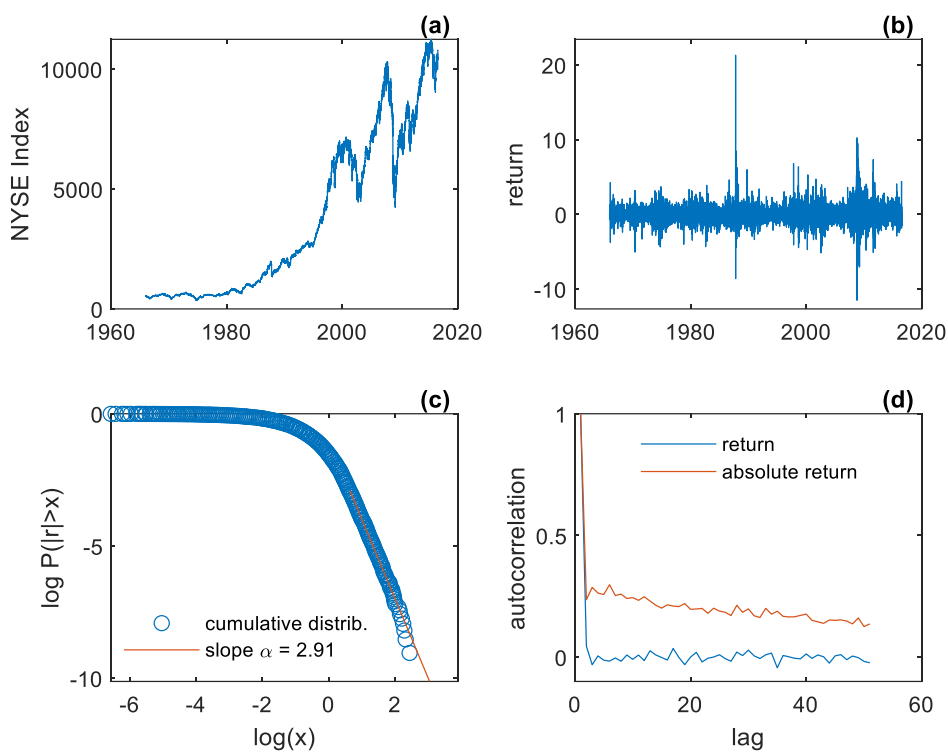

**Figure 2.** NYSE composite daily index.

## 3. The model

Following a traditional dichotomy, consider a financial market populated by two types of traders: (short-term) speculators, who buy an asset for anticipated capital gains; and (long-run fundamental-value) investors, who buy an asset based on its fundamental value. Let the (excess) demands of an investor and a speculator be respectively[8]:

$$Z_{it} = \mu \frac{V_{it}^e - P_t}{P_t}, \tag{1}$$

$$Z_{st} = \gamma \frac{P_{st}^e - P_t}{P_t}, \tag{2}$$

where $P_{st}^e$ is a speculator's estimation of the asset's future price, $V_{it}^e$ is an investor's estimation the asset's present value, and the parameters $\mu, \gamma > 0$. Let $M_t$ and $N_t$ be respectively the numbers of investors and speculators active in period $t$. The overall (market) excess demand is

$$Z_t = \mu M_t \frac{V_t^e - P_t}{P_t} + \gamma N_t \frac{P_t^e - P_t}{P_t}, \tag{3}$$

where $P_t^e = N_t^{-1} \sum_s P_{st}^e$ and $V_t^e = M_t^{-1} \sum_i V_{it}^e$, namely, the average investor valuation (hereafter referred to simply as 'the value' of the security) and the average speculator anticipated future price.

Assume the following standard price adjustment, in accordance with the market microstructure literature[9]:

---

[8] Because nonlinearity adds no further insight to this theory, we assume these standard linear supply and demand functions, which can be viewed as first-order linear approximations of more general functions; also, since financial supply and demand can be treated symmetrically (by treating supply formally as a negative demand), one can think directly in terms of a trader's excess demand, which is a demand or a supply, depending on the sign.

[9] The seminal work is Kyle (1985). A distinction may be in order here: a 'price adjustment function' models the overall price impact of the competition of buyers and sellers in a market; empirical

$$r_t = \beta \frac{Z_t}{L_t}, \tag{4}$$

where $L_t$ is the overall market liquidity (or market depth) and $\beta > 0$. Let the overall price impact of speculative and investment orders be denoted respectively as

$$m_t = \beta\mu M_t/L_t, \tag{5}$$

$$n_t = \beta\gamma N_t/L_t. \tag{6}$$

The two equations (3) and (4) combined yield:

$$r_t = m_t \frac{V_t^e - P_t}{P_t} + n_t \frac{P_t^e - P_t}{P_t}, \tag{7}$$

*A purely news-driven investment market model*

Let the arrival of exogenous news relevant to investors and speculators be modeled as random events $I_t$ and $J_t$ occurring with probability $\mathbb{P}(I_t)$ and $\mathbb{P}(J_t)$ and leading the traders to revise additively their prior estimations of the asset by the amounts $\varepsilon_t$ and $\nu_t$ respectively (which can be assumed normally distributed by aggregation). (There is no harm in assuming $I_t = J_t$, namely a common access to the same news by all the traders.) Thus the traders' valuations of the asset follow a random walk: $V_t^e = V_{t-1}^e + \varepsilon_t \mathbf{1}(I_t)$ and $P_t^e = P_{t-1}^e + \nu_t \mathbf{1}(I_t)$, where $\mathbf{1}(I_t)$ and $\mathbf{1}(J_t)$ are the indicator functions associated with the advent of the news. The amount of information the news reveal to the traders about the asset's worth can be precisely quantified: $-\log(\mathbb{P}(I_t))$ and $-\log(\mathbb{P}(I_t))$, respectively. This *news*-driven random-walk of traders' expectations should be distinguished from a standard assumption in the agent-based literature, introduced perhaps by Lux and Marchesi (1999), whereby an asset's fundamental value is modeled as a *noise*-driven random walk, where the noise is a Gaussian white noise. The difference between a news and a noise is simple: a noise can be formally defined as a news that carries zero information ($\mathbb{P}(I_t) = 1$, $\mathbb{P}(J_t) = 1$). [10]

---

evidence suggests it is linear in financial markets (Cont et al., 2014); a related but different concept is the 'price impact' of a trade or series of trades, which is typically a concave function of trade volume (Bouchaud, 2010). The first concept is relevant for a theorist studying a market as a whole; the second, perhaps for a trader wishing to minimize the execution cost of a given trade volume.

[10] We are grateful to a reviewer whose comment makes us aware of the need to emphasize explicitly the basic difference between a news and a noise. The reviewer suggests also that the noise-driven random walk of fundamental value, along with traders' heterogeneity, may be responsible for clustered volatility in some agent-based models such as that by He and Li (2012), although these models put forward a complex mix of nonlinear mechanisms. Yet most agent-based models who assume the same Gaussian *white noise* in the fundamental-value dynamics, starting from Lux and Marchesi (1999), make the opposite claim, showing that the noise-driven random walk in their model has nothing to do with the stylized facts, and in fact they assume the Gaussian white noise precisely so that none of the emergent financial stylized facts in their models be attributed to this white noise: "In order to ensure that none of the typical characteristics of financial prices can be traced back to exogenous factors, we assume that the relative changes of [fundamental value] are Gaussian random variables." (Lux & Marchesi, 1999, p. 499). In fact, Lux and Marchesi (2000) showed that the fat-tail and volatility clustering in their model hold even when the fundamental value is constant. This is the case in this paper's model as well, as emphased below (Figure 7).

All in all, the asset's price dynamics reads:

$$P_t = (1 + r_t)P_{t-1}, \tag{8}$$

$$r_t = m_t \frac{V_t^e - P_t}{P_t} + n_t \frac{P_t^e - P_t}{P_t}, \tag{9}$$

$$V_t^e = V_{t-1}^e + \varepsilon_t \mathbf{1}(I_t), \tag{10}$$

$$P_t^e = P_{t-1}^e + \nu_t \mathbf{1}(J_t). \tag{11}$$

Because both types of traders are behaviorally equivalent, the investor-speculator dichotomy is of no substance in this specific model: $P^e$ is equivalent to $V^e$, and both are entirely driven by exogenous news. In other words, the market thus modeled is in fact a non-speculative purely news-driven market of investors. Figure 3 shows a simulation of this model, and Figure 4 shows 5 superposed sample paths of the model. (All the parameter specifications are reported in Table 1 in the discussions section.) The clustering of volatility is generic in this model; but, as is clear from Figure 3, the model suffers from an obvious non-stationarity of the return process due to the double random walk of the traders' expectations; thus the graphical impression of a robust fat tail is an artefact: it does not make sense to say that the distribution is fat-tailed (since the returns are in fact drawn from different distributions: for example, the standard deviation of the return varies greatly from sample to sample, as should be expected).

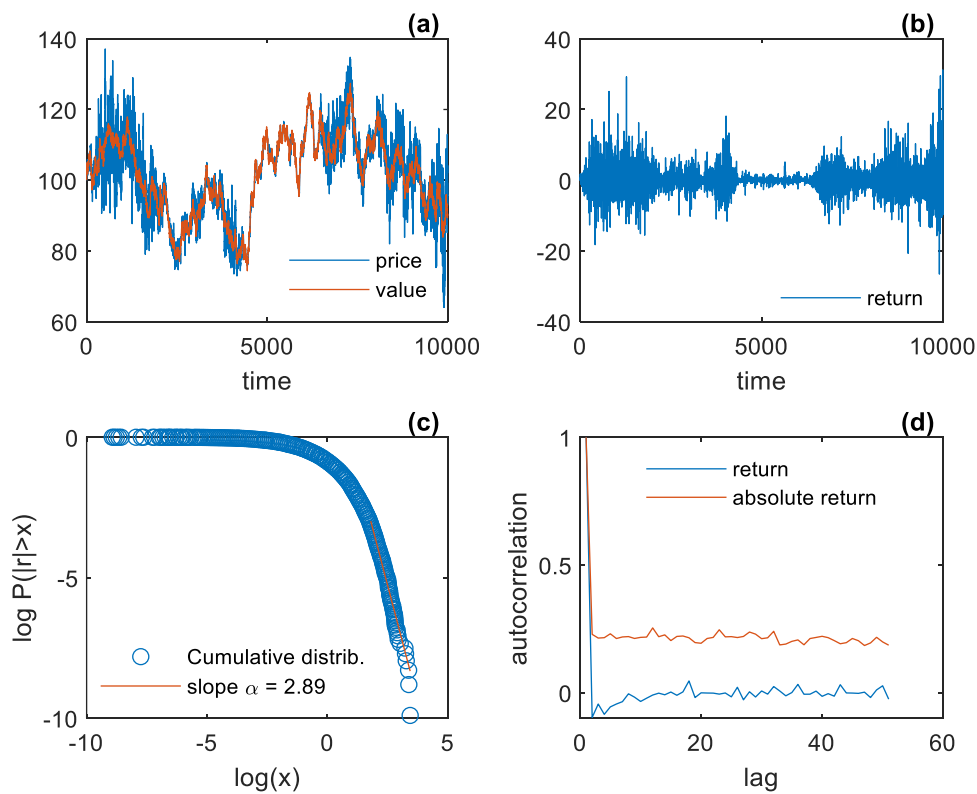

**Figure 3.** A purely news-driven market model.

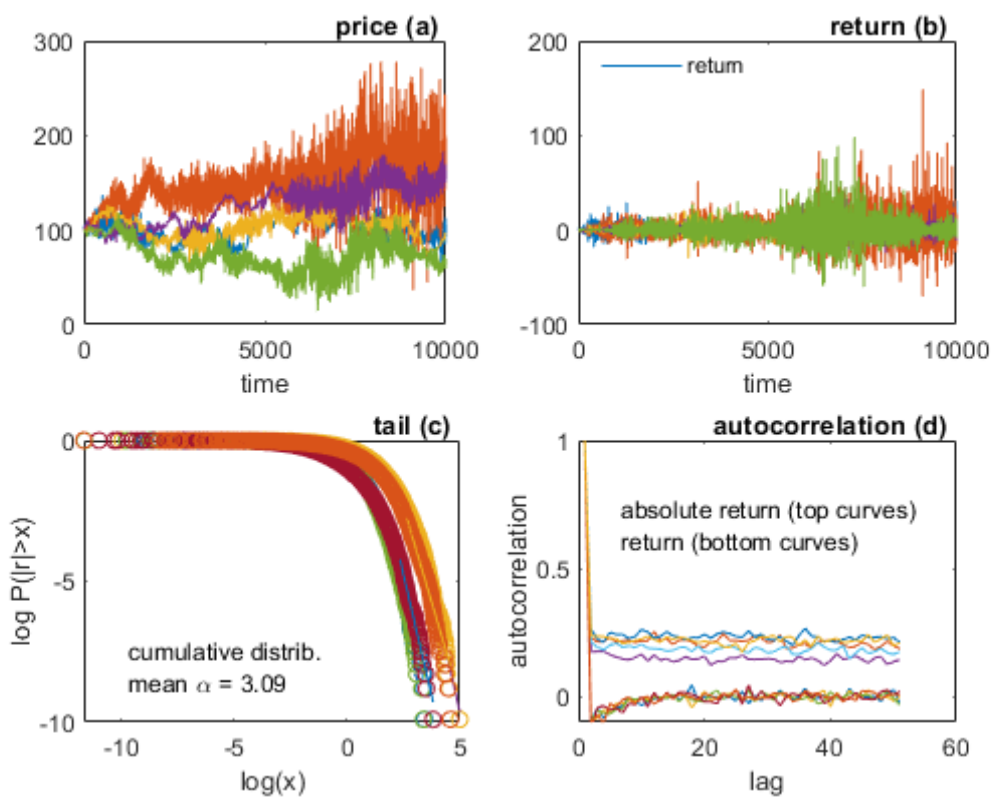

**Figure 4.** The purely news-driven market model: 5 simulations superposed.

*A purely speculative trend-following market model*

Let the speculators' anticipated return be denoted

$$r_t^e \equiv \frac{P_t^e - P_t}{P_t}. \tag{12}$$

Trend-following implies that speculators' overall anticipated return is of the form $r_t^e = \sum_{h=1}^{H} \omega_{ht} r_{t-k} + \nu_t \mathbf{1}(J_t)$, where we have added the impact of exogenous news on speculators' expectations. The weighting scheme $\{\omega_{ht}\}$ can be computed explicitly from standard moving-average trend-following techniques used by financial practitioners (Beekhuizen & Hallerbach, 2017). In a purely speculative market ($m_t = 0$), the asset's return is then $r_t = n_t \sum_{h=1}^{H} \omega_{ht} r_{t-h} + n_t \nu_t \mathbf{1}(J_t)$. This RCAR model generates, under quite general and mild technical conditions, a strictly stationary power law tail $\mathbb{P}(|r_t|>x) \sim C x^{-\alpha}$, where the exponent $\alpha$ depends solely on the joint distribution of $(n_t, \omega_{ht})$, and not on the exogenous news (Klüppelberg & Pergamenchtchikov, 2004; Buraczewski et al., 2016). But this strict stationary comes at a price, as noted in the introduction: for any such autoregressive model, and for any arbitrary function $f$, $\text{cov}[f(r_t), f(r_{t+h})]$, when it is well-defined, decays rapidly, at an exponential rate, with the lag $h$ (Mikosch & Starica, 2000; Basrak et al., 2002). So volatility, whether measured as $|r_t|$, $r^2$, or more generally by any function $f$, cannot be long-range correlated in this purely speculative trend-following model.

*A more general model*

The two polar models emphasize a tension between the endogenous and the exogenous causes of volatility: the purely exogenous, news driven-driven, expectations, produces a clustering of volatility but induces a trivial non-stationarity; whereas the purely endogenous feedback-inducing trend-following expectations generates a stationary power-law tail but cannot account for volatility clustering. The simplest compromise between these two notions consists of maintaining a purely exogenous, news-driven investors' valuations, but to assume that the speculators' anticipated return is partly endogenous (self-referential, or reflexive) and to write $r_t^e = ar_{t-1}^e + \nu_t \mathbf{1}(J_t)$, where $0 < a < 1$, to capture the (exponentially decaying) short-memory of speculators' concerning a fundamental news, but $a \approx 1$, so that incoming news have a lasting enough impact upon the speculators' expectations.[11] The purely news-driven random walk of investors' valuations, on the other hand, implies that the asset's value incorporates all the fundamental news (news relevant to investors) in the sense that $V_t^e = V_0^e + \sum_{k=1}^t \nu_k \mathbf{1}(I_k)$, making $V_t^e$ the natural estimate of the asset's fundamental value in this model.

All in all, the asset's price dynamics in the general model reads:

$$P_t = (1 + r_t) P_{t-1}, \tag{13}$$

$$r_t = n_t r_t^e + m_t \frac{V_t^e - P_t}{P_t}, \tag{14}$$

$$V_t^e = V_{t-1}^e + \nu_t \mathbf{1}(I_t), \tag{15}$$

$$r_t^e = ar_{t-1}^e + \varepsilon_t \mathbf{1}(J_t), \tag{16}$$

Figures 3-6 are simulations of the model using the parameter specifications in Table 1.

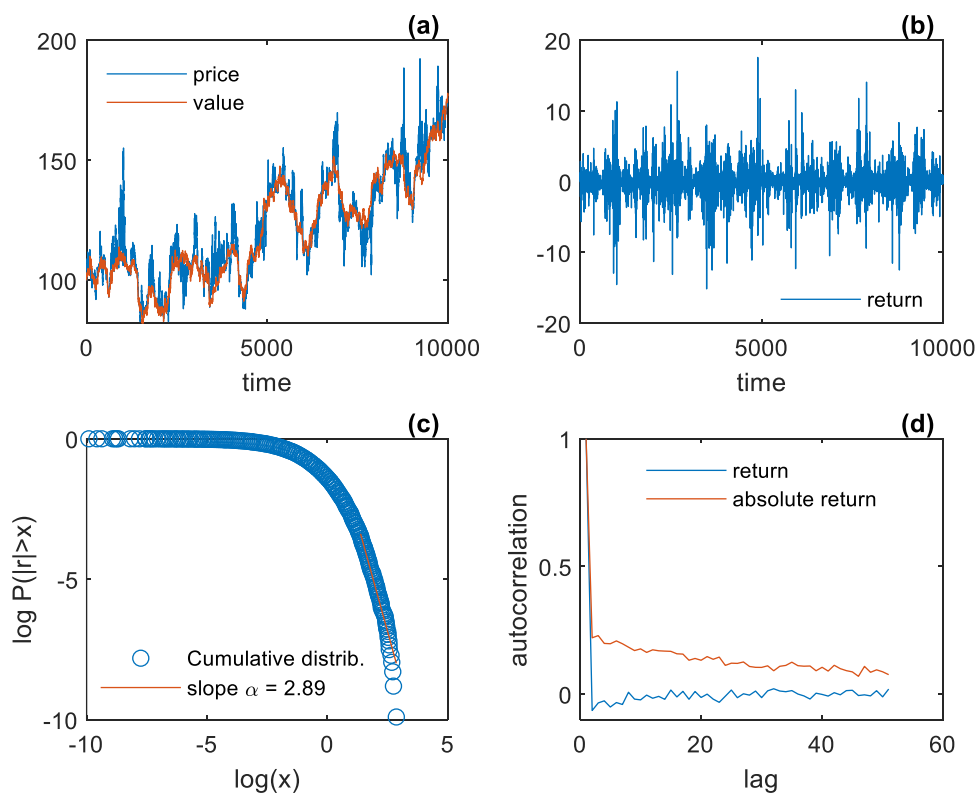

**Figure 5.** The general model: illustration 1.

---

[11] An arbitrarily general AR model is recently suggested by Shi, Luo, and Li (2019), which replicates and studies in detail the robustness of an earlier working version of this paper's model (titled 'The random walk behind volatility clustering, 2016'). But the choice $a \approx 1$ is crucial for volatility clustering.

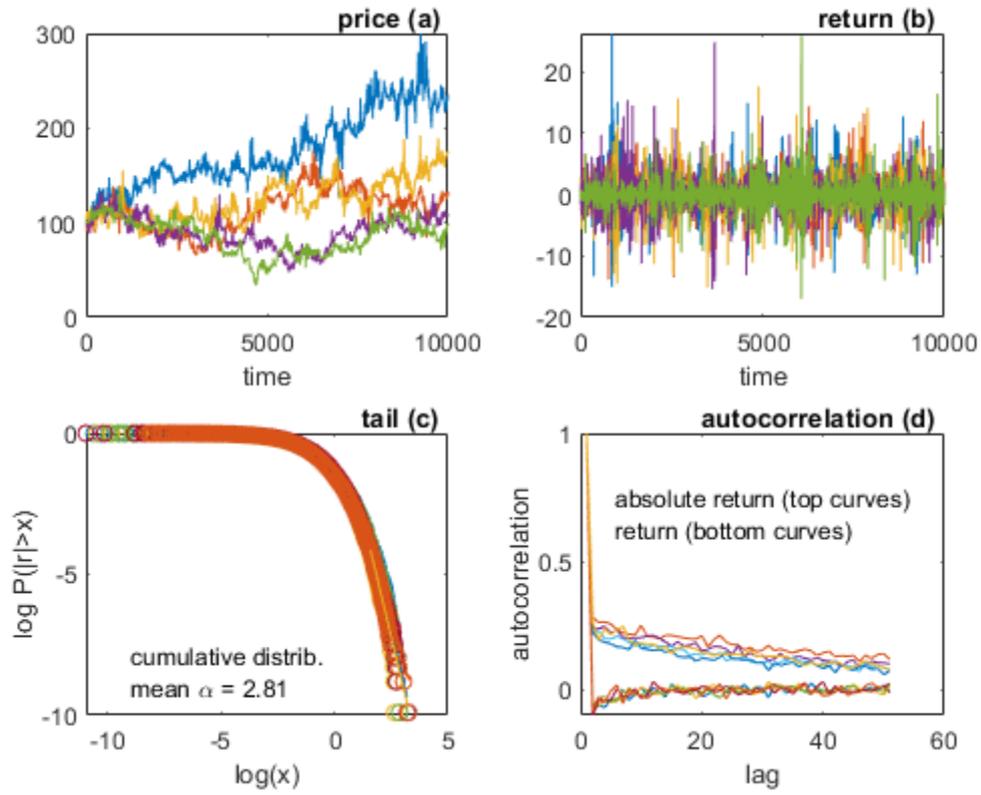

**Figure 6.** The general model: 5 simulations superposed.

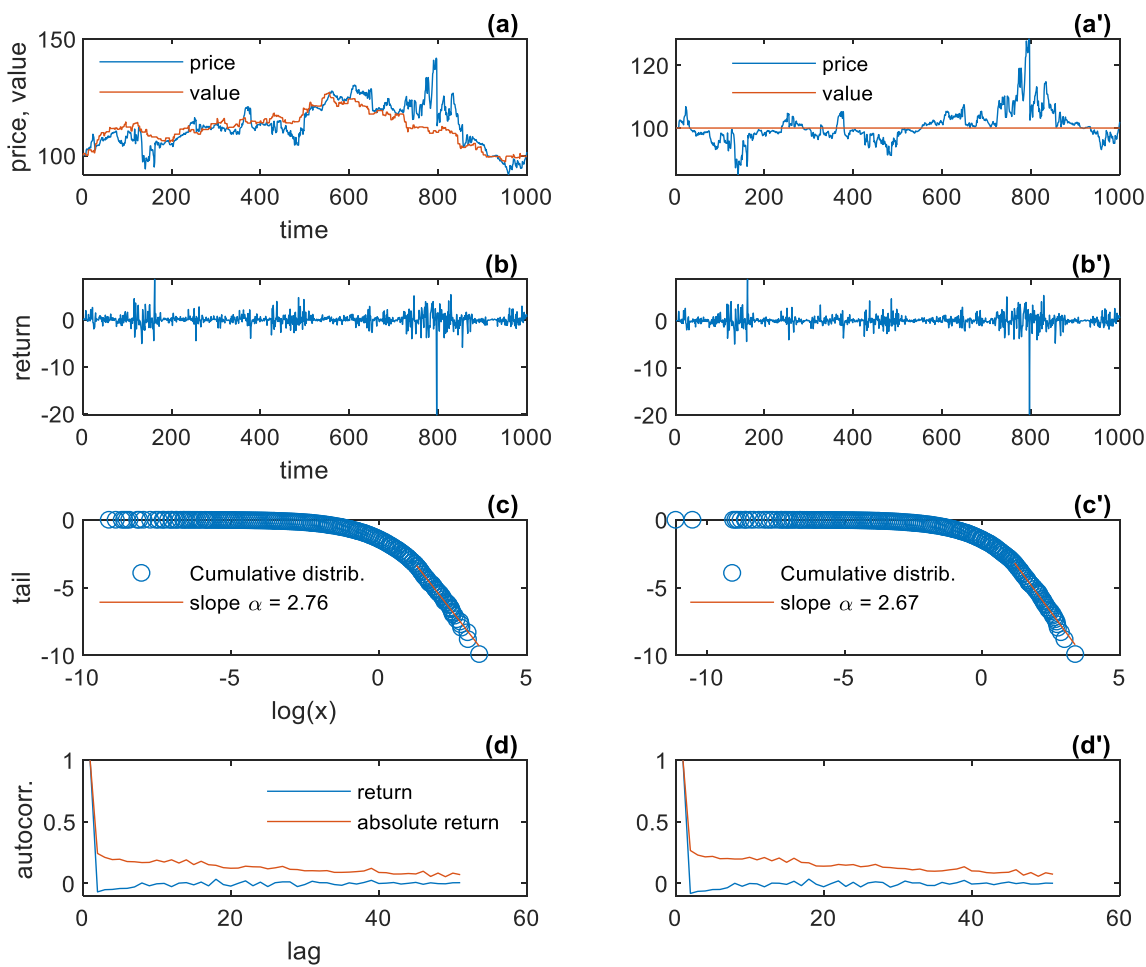

**Figure 7.** The general model: variable versus constant fundamental value.[12]

## 4. Discussion

Both the fat tail and the volatility clustering are generic and robust in the model: they hold for a broad class of distributions and parameters: the specifications in Figure 3-7 are chosen merely for illustration, except to reflect realistic orders of magnitude compared to empirical data (notably the standard deviation of return which is typically around 1). Also, in all the simulations: $T = 10000$ periods; $P_0 = V_0^e = P^e = 100$, $r_1 = r_1^e = 0$; $\{n_t\}$ and $\{m_t\}$ are exponentially distributed iid processes; $\{\varepsilon_t\}$ and $\{\nu_t\}$ are zero-mean Gaussian iid processes. The differing parameter choices are reported in Table 1.

**Table 1.** Parameter specifications.

|  | *Figure 1* GE | *Figure 2* NYSE | *Figure 3* Model 1 | *Figure 4*[1] Model 1 | *Figure 5* Model 2 | *Figure 6*[1] Model 2 | *Figure 7* Model 2 |
|---|---|---|---|---|---|---|---|
| *Parameters* | | | | | | | |
| mean ($m$) | | | 0.1 | 0.1 | 0.2 | 0.2 | 0.2 |
| mean ($n$) | | | 0.1 | 0.1 | 0.1 | 0.1 | 0.1 |
| std($\varepsilon$) | | | 1 | 1 | 1 | 1 | 1 |
| std($\nu$) | | | 1 | 1 | 0.04 | 0.04 | 0.04 |

---

[12] For greater visibility, only the first 1000 periods out of 10000 are shown in the subplots (a), (a'), (b), and (b').

| | | | | | | | |
|---|---|---|---|---|---|---|---|
| prob(News I) | | | 0.5 | 0.5 | 0.3 | 0.3 | $0.3^3$; $0^4$ |
| prob(News J) | | | 0.5 | 0.5 | 0.1 | 0.1 | 0.1 |
| *a* (*feedback*) | | | | | 0.99 | 0.99 | 0.99 |
| mean(*r*) | -0.04 | -0.02 | 0.04 | $0.16^2$ | 0.01 | $0.01^2$ | $0.01^3$; $0.01^3$ |
| std(*r*) | 1.62 | 1.01 | 2.68 | $4.99^2$ | 1.43 | $1.46^2$ | $1.48^3$; $1.44^3$ |

[1]Five sample paths superposed. [2]Average over 5 sample paths. [3]Left subplots (a)-(d). [4]Right subplots (a')-(d').

## 5. Conclusion

This paper suggests a simple explanation for excess and clustered volatility in financial markets through a simple synthesis of the two major paradigms in financial theory. Excess volatility means that that price fluctuations are too high given the underlying fundamentals, which is an intrinsic feature of the model, owing to the amplifying feedback intrinsic to speculative trading, as illustrated more strikingly in Figure 7, in which the fundamental value is kept constant. Clustered volatility simply reflects, in this theory, the traders' persistent memory of exogenous news concerning the asset's present value or future price. The two empirical facts are thus reduced to simple explanations, through basic *linear processes* that capture investors' long-memory versus speculators' shorter-memory of real news.

**Acknowledgments:**

For helpful comments and suggestions, the author thanks V. L. Smith, C. Wihlborg, D. Porter, and two anonymous reviewers.